\documentstyle[multicol,aps,psfig]{revtex}
\renewcommand{\narrowtext}{\begin{multicols}{2}
\global\columnwidth20.5pc\noindent}
\renewcommand{\widetext}{\end{multicols}
\global\columnwidth42.5pc}
\begin{document}
\draft
\preprint{6 February 2001}
\title{Quantum and Thermal Phase Transitions of Halogen-Bridged
       Binuclear Transition-Metal Complexes}
\author{Shoji Yamamoto}
\address
{Department of Physics, Okayama University,
 Tsushima, Okayama 700-8530, Japan}
\date{Received 6 February 2001}
\maketitle
\begin{abstract}
Aiming to settle the controversial observations for halogen-bridged
binuclear transition-metal (MMX) complexes, finite-temperature
Hartree-Fock calculations are performed for a relevant two-band
Peierls-Hubbard model.
Thermal, as well as quantum, phase transitions are investigated with
particular emphasis on the competition between electron itinerancy,
electron-phonon interaction and electron-electron correlation.
Recently observed distinct thermal behaviors of two typical MMX
compounds
Pt$_2$(CH$_3$CS$_2$)$_4$I and
(NH$_4$)$_4$[Pt$_2$(P$_2$O$_5$H$_2$)$_4$I]$\cdot$2H$_2$O
are supported and further tuning of their electronic states is
predicted.
\end{abstract}
\narrowtext

   There is a class of quasi-one-dimensional materials containing
chains of transition-metal (M) complexes bridged by halogens (X).
Their representatives such as
Wolffram's red ($\mbox{M}=\mbox{Pt},\mbox{X}=\mbox{Cl}$) and
Reihlen's green ($\mbox{M}=\mbox{Pt},\mbox{X}=\mbox{Br}$) salts
\cite{RC095} are composed of mononuclear metal complexes and are
thus referred to as MX chain compounds.
These PtX chains generally exhibit a mixed-valence ground state with
strong dimerization of the X sublattice.
When Pt is replaced by Ni \cite{HT261,KT341}, a mono-valence
regular-chain structure is instead stabilized, which is nothing but
a tuning from the Peierls to Mott insulator.
The electronic state of the MX chain system can such widely be tuned
that the optical energy gap changes from $0.8$ to $3.3$ eV by varying
the transition metals, the surrounding ligands, the bridging halogens
and the counter ions \cite{HO009}.
In recent years, further efforts have been devoted to bridging
polynuclear metal complexes by halogens and physical as well as
chemical interest in this system has been renewed.
A new group, MMX, consists of two families,
R$_4$[Pt$_2$(pop)$_4$X]$\cdot$$n$H$_2$O
($\mbox{X}=\mbox{Cl},\mbox{Br},\mbox{I}$;
 $\mbox{R}=\mbox{Li},\mbox{K},\mbox{Cs},\mbox{NH}_4$;
 $\mbox{pop}=\mbox{diphosphonate}
 =\mbox{H}_2\mbox{P}_2\mbox{O}_5^{\,2-}$)
\cite{PS489,MC606,MK420,RC409} and
M$_2$(dta)$_4$I
($\mbox{M}=\mbox{Pt},\mbox{Ni}$;
 $\mbox{dta}=\mbox{dithioacetate}=\mbox{CH}_3\mbox{CS}_2^{\,-}$)
\cite{CB444,CB815}, which are generically called the pop and dta
complexes, respectively.
The MMX chain system potentially provides not a few topics of
qualitatively new aspect.
The formal oxidation state of the metal ions is $3+$ in MX chains,
whereas $2.5+$ in MMX chains.
Therefore, MMX chains possess an unpaired electron per metal dimer
even in their trapped-valence states, which is in contrast with
the valence-trapped state consisting of M$^{2+}$ and M$^{4+}$ in MX
chains.
The M($d_{z^2}$)-M($d_{z^2}$) direct overlap in MMX chains
effectively reduces the on-site Coulomb repulsion due to its
$d_{\sigma^*}$ character and may therefore enhance the electrical
conductivity.
In the dta complexes with no counter ion, the chains weakly interact
with each other through van der Waals contacts but are free from
relatively strong hydrogen bonds, raising an increased possibility of
the metal sublattice, as well as the halogen sublattice, being
distorted.
Nowadays, the chemical explorations along this line have reached a
novel MMMMX chain compound
HH-[Pt(2.5+)$_4$(NH$_3$)$_8$($\mu$-C$_4$H$_6$NO)$
_4$Cl](ClO$_4$)$_3$Cl$_2$ \cite{KS366}.

   In response to recent stimulative observations for MMX chain
compounds, we here study quantum and thermal phase competitions in
one-dimensional unit-assembled spin-charge-lattice coupling systems.
We employ the $\frac{5}{6}$-filled one-dimensional two-band
three-orbital extended Peierls-Hubbard Hamiltonian:
\begin{eqnarray}
   {\cal H}
   &=&\sum_{m,s}\sum_{j=1,2}
      \big(\varepsilon_{\rm M}-\beta l_{j:n}\big)n_{j:m,s}
    + \sum_{m,s}
      \varepsilon_{\rm X}n_{3:m,s}
      \nonumber \\
   &-&\sum_{m,s}\sum_{j=1,2}
      \big(t_{\rm MX}-\alpha l_{j:n}\big)
      \big(
       a_{j:n,s}^\dagger a_{3:n,s}+a_{3:n,s}^\dagger a_{j:n,s}
      \big)
      \nonumber \\
   &-&\sum_{m,s}
      t_{\rm MM}
      \big(
       a_{1:n,s}^\dagger a_{2:n-1,s}
      +a_{2:n-1,s}^\dagger a_{1:n,s}
      \big)
      \nonumber \\
   &+&\sum_{m}\sum_{j=1,2}
      U_{\rm M}\,n_{j:m,+}n_{1:m,-}
    + \sum_{m}
      U_{\rm X}\,n_{3:n,+}n_{3:n,-}
      \nonumber \\
   &+&\sum_{m,s,s'}\sum_{j=1,2}
      V_{\rm MX}\,n_{j:m,s}n_{3:m  ,s'}
      \nonumber \\
   &+&\sum_{m,s,s'}
      V_{\rm MM}\,n_{1:m,s}n_{2:m-1,s'}
    + \sum_{m}\sum_{j=1,2}
      \frac{K}{2}l_{j:m}^2\,,
   \label{E:H}
\end{eqnarray}
where $n_{j:m,s}=a_{j:m,s}^\dagger a_{j:m,s}$ with
$a_{j:m,s}^\dagger$ being the creation operator of an electron with
spin $s=\pm$ (up and down) for the M $d_{z^2}$ ($j=1,2$) or X $p_z$
($j=3$) orbital in the $m$th MXM unit, and
$l_{j:m}=(-1)^j(u_{j:m}-u_{3:m})$ with $u_{j:m}$ being the
chain-direction displacement of the metal ($j=1,2$) or halogen
($j=3$) in the $m$th MXM unit from its equilibrium position.
$\alpha$ and $\beta$ are, respectively, the intersite and intrasite
electron-lattice coupling constants, while $K$ is the metal-halogen
spring constant.
We assume, based on the thus-far reported experimental observations,
that every M$_2$ moiety is not deformed, namely, $u_{1:n}=u_{2:n-1}$.
$\varepsilon_{\rm M}$ and $\varepsilon_{\rm X}$ are the on-site
energies of isolated metal and halogen atoms, respectively.
The electron hoppings between these levels are modeled by
$t_{\rm MM}$ and $t_{\rm MX}$, whereas the electron-electron Coulomb
interactions by $U_{\rm M}$, $U_{\rm X}$, $V_{\rm MM}$ and
$V_{\rm MX}$.
Though different-site Coulomb interactions may in principle be
coupled with lattice displacements in a two-band description, any
alternation of them is usually neglected renormalizing $\beta$ under
the mean-field treatment of $\sum_s n_{3:n,s}$.
We always set $t_{\rm MX}$ and $K$ both equal to unity.

\begin{figure}
\begin{flushleft}
\mbox{\qquad\qquad\qquad\qquad\ 
      \psfig{figure=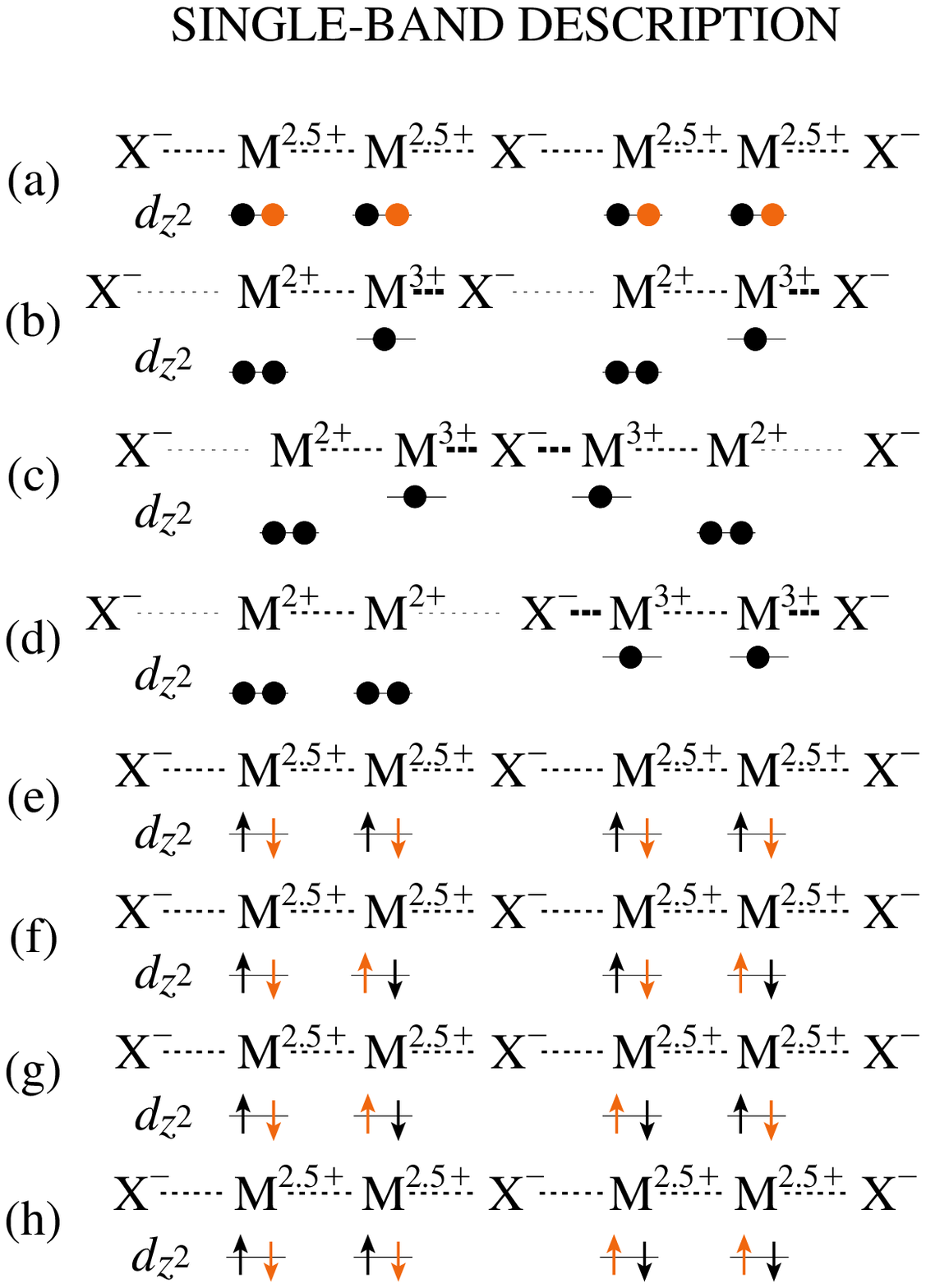,width=72mm,angle=0}
      $\!\!\!\!\!\!\!\!\!\!\!\!$
      \psfig{figure=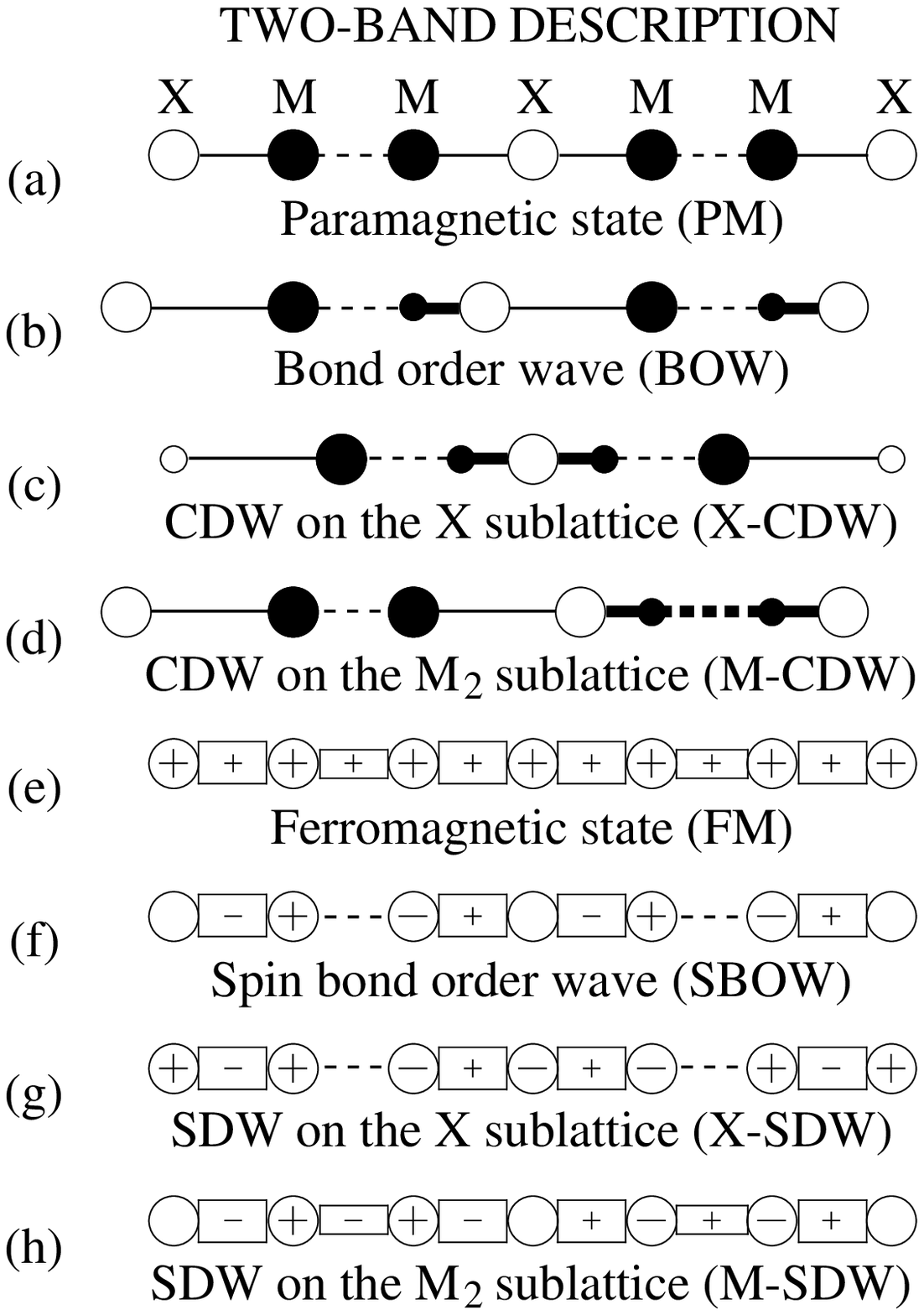,width=72mm,angle=0}}
\end{flushleft}
\vskip 0mm
\caption{Schematic representation of possible density-wave states
         within the single-band approximation and in terms of the
         two-band description.
         In the left half, the circles and arrows denote electron
         total (when closed) or half (when shaded) occupancy with
         unpolarized and polarized spins, respectively, where the
         valence numbers denote formal oxidation states, namely,
         $2+$ and $3+$ should generally be regarded as $(2+\delta)+$
         and $(3-\delta)+$ ($0\leq\delta\leq 0.5$), respectively.
         In the right half, the various circles and segments
         qualitatively denote the variation of local electron
         densities and bond orders, respectively, whereas the signs
         $\pm$ in circles and strips denote the alternation of local
         spin densities and spin bond orders, respectively, where the
         bond and spin bond orders between site $i$ at the $n$th MXM
         unit and site $j$ at the $m$th MXM unit are, respectively,
         defined as
         $p_{i:n;j:m}=\sum_s a_{i:n,s}^\dagger a_{j:m,s}$ and
         $t_{i:n;j:m}
         =\frac{1}{2}\sum_s s a_{i:n,s}^\dagger a_{j:m,s}$.
         Symbols shifted from the regular position qualitatively
         represent lattice distortion.}
\label{F:DW}
\end{figure}

   Current empirical arguments on the electronic structures of MMX
compounds are aiming to distinguish between four types of
one-dimensional charge ordering mode \cite{MK}, which are
illustrated as (a) to (d) in Fig. \ref{F:DW}, assuming that the X
$p_z$ orbitals are stably filled and irrelevant.
Quite recently, potential magnetic phases have also been pointed out
((e) to (h) in Fig. \ref{F:DW}) and a systematic study of
broken-symmetry solutions in the two-band scheme has been presented
\cite{SY183}.
The two-band description of all the phases is also given in Fig.
\ref{F:DW}.
We learn that (c) and (g) should be characterized by charge and spin
modulation on the X sublattice, respectively, rather than by any
density wave on the M$_2$ sublattice.
The relevance of the X $p_z$ orbitals in these states has indeed been
visualized numerically \cite{SY}.

   Most of the ground states of the pop-family MMX compounds have now
been assigned to (d) \cite{LB155,SJ415,NK040,YW195}, though early
structural analyses \cite{MK420,RC409} were rather controversial.
However, due to the relatively small Peierls gaps in comparison with
those of MX compounds, they can be tuned from the strongly
valence-trapped state toward the paramagnetic or antiferromagnetic
valence-delocalized state by replacing halogen and/or counter ions
\cite{LB155,MY321}.
Since the halogen character increases in the $d_\sigma$ Pt-Pt wave
functions in the order $\mbox{Cl}<\mbox{Br}<\mbox{I}$, even further
charge fluctuation toward (c) \cite{SC723} may be observed in the
iodo complexes.
Such a reverse Peierls instability can be reinforced under external
pressure \cite{BS405,MS066}.
In general, the complete valence localization breaks down with
increasing temperature \cite{LB155,NK040}.
(NH$_4$)$_4$[Pt$_2$(pop)$_4$I]$\cdot$$2$H$_2$O clearly exhibits a
thermal phase transition from (d) to (a) \cite{MY321}.

   There has also been a controversy over the electronic structures
of the dta-family compounds.
Early investigations \cite{CB444,CB815} reported similar
semiconducting behaviors for both Pt and Ni complexes.
However, later magnetic-susceptibility \cite{MY207} and proton-spin
relaxation-time \cite{RI907} measurements claimed to distinguish
Ni$_2$(dta)$_4$I from Pt$_2$(dta)$_4$I elucidating the distinct
antiferromagnetic spin structure with strong exchange coupling for
the former.
Furthermore, recent extensive physical measurements \cite{HK068} 
on Pt$_2$(dta)$_4$I have revealed its metallic conduction above room
temperature as well as a totally different scenario for its valence
structure:
With decreasing temperature, there occurs a metal-semiconductor
transition at room temperature, which can be regarded as the
transition from the valence-delocalized state (a) to the
trapped-valence state (b), and further transition to the
Peierls-insulating charge-ordering mode (c) follows around
$80\,\mbox{K}$.

\widetext
\begin{figure}
\begin{flushleft}
\mbox{\psfig{figure=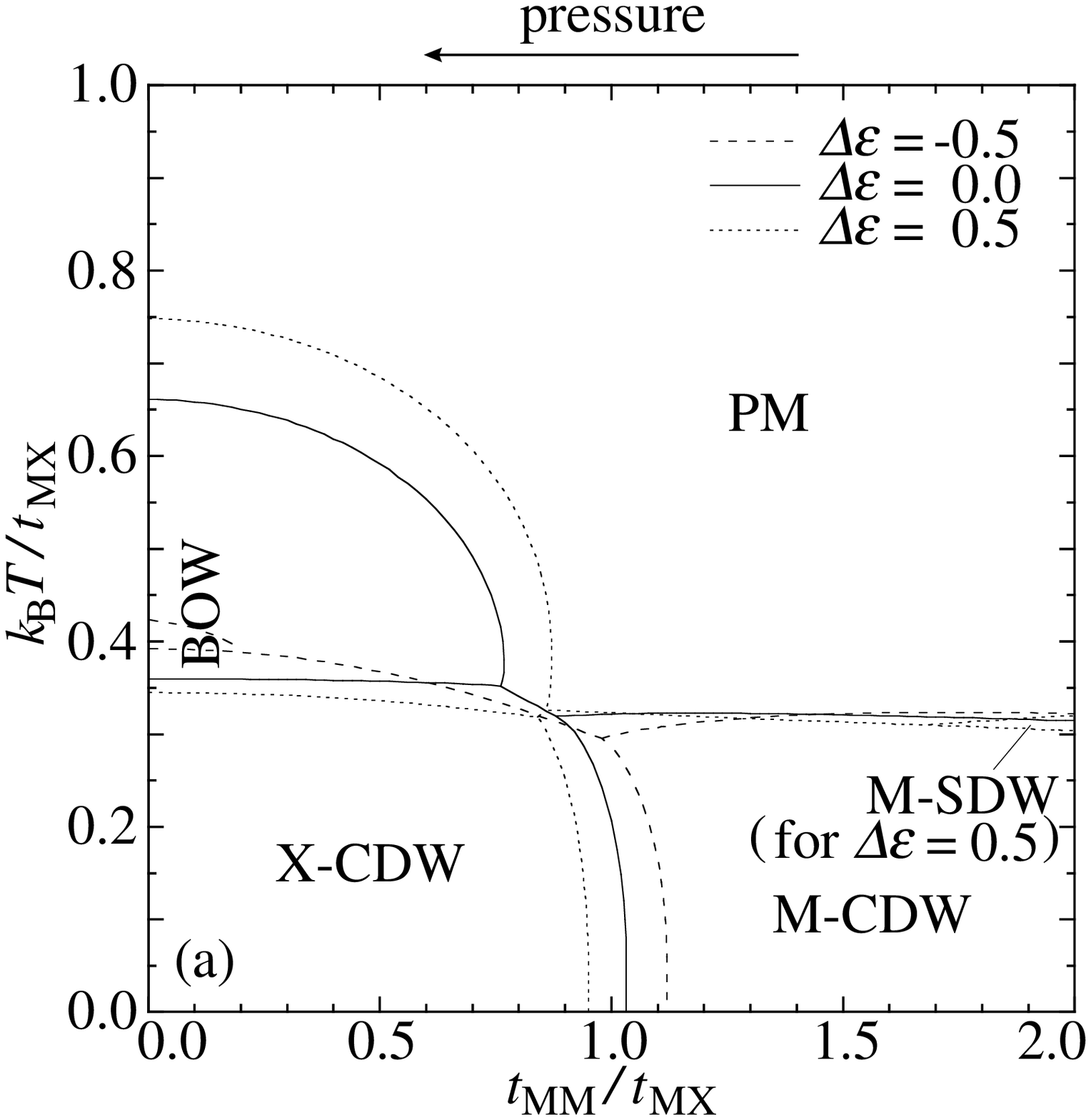,width=49.0mm,angle=0}$\!\!\!\!$
      \psfig{figure=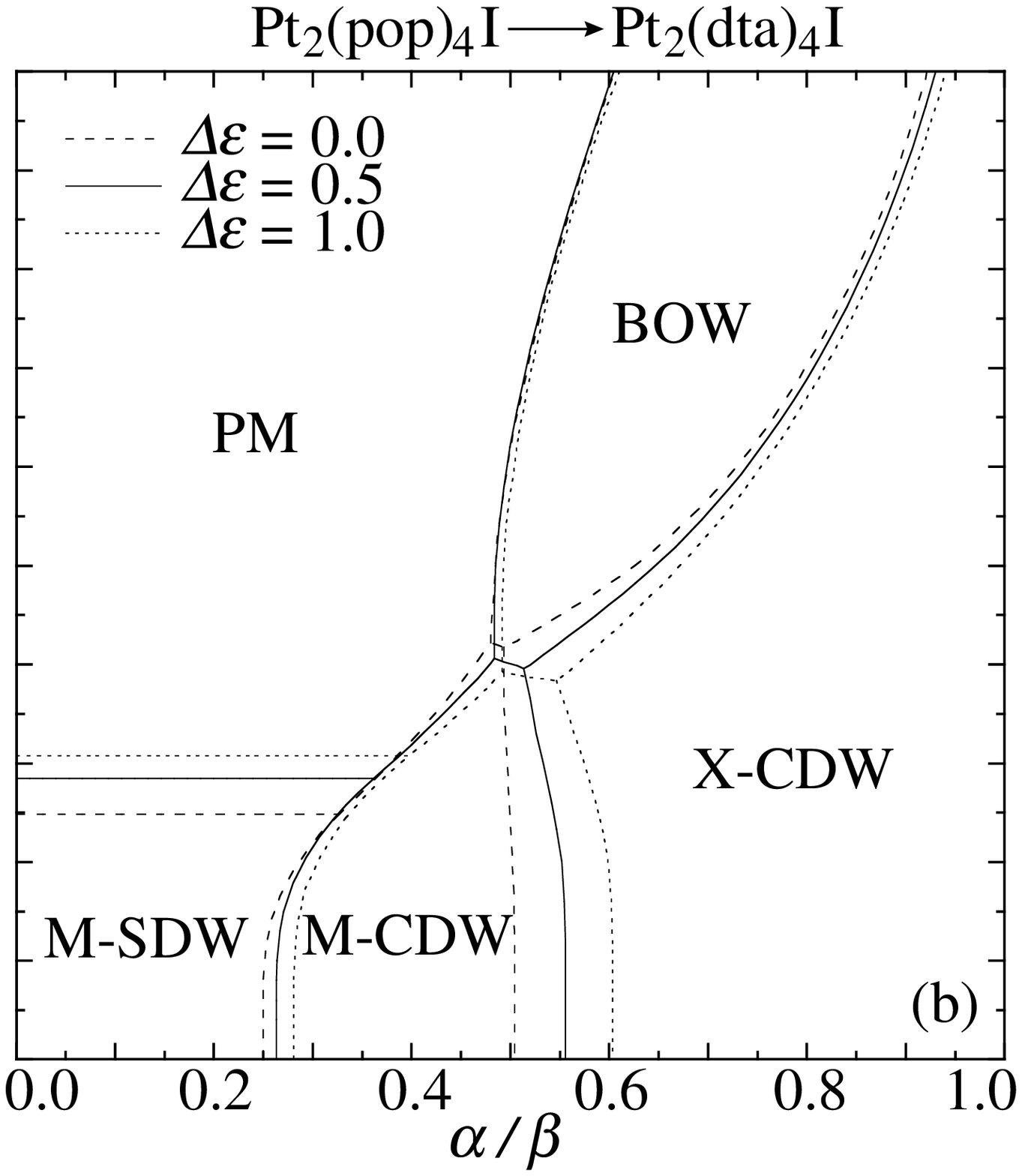,width=43.1mm,angle=0}$\!\!\!\!$
      \psfig{figure=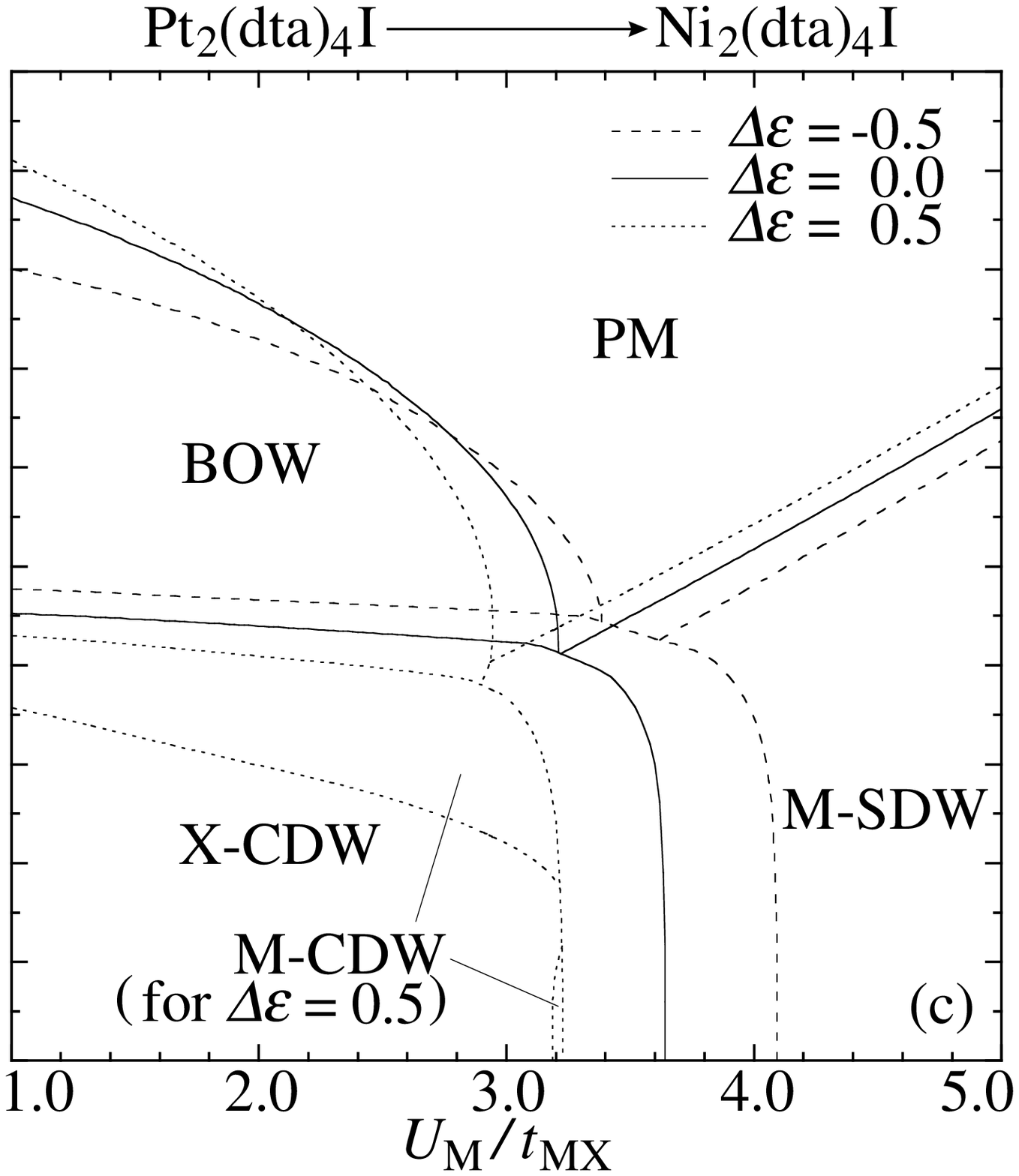,width=43.1mm,angle=0}$\!\!\!\!$
      \psfig{figure=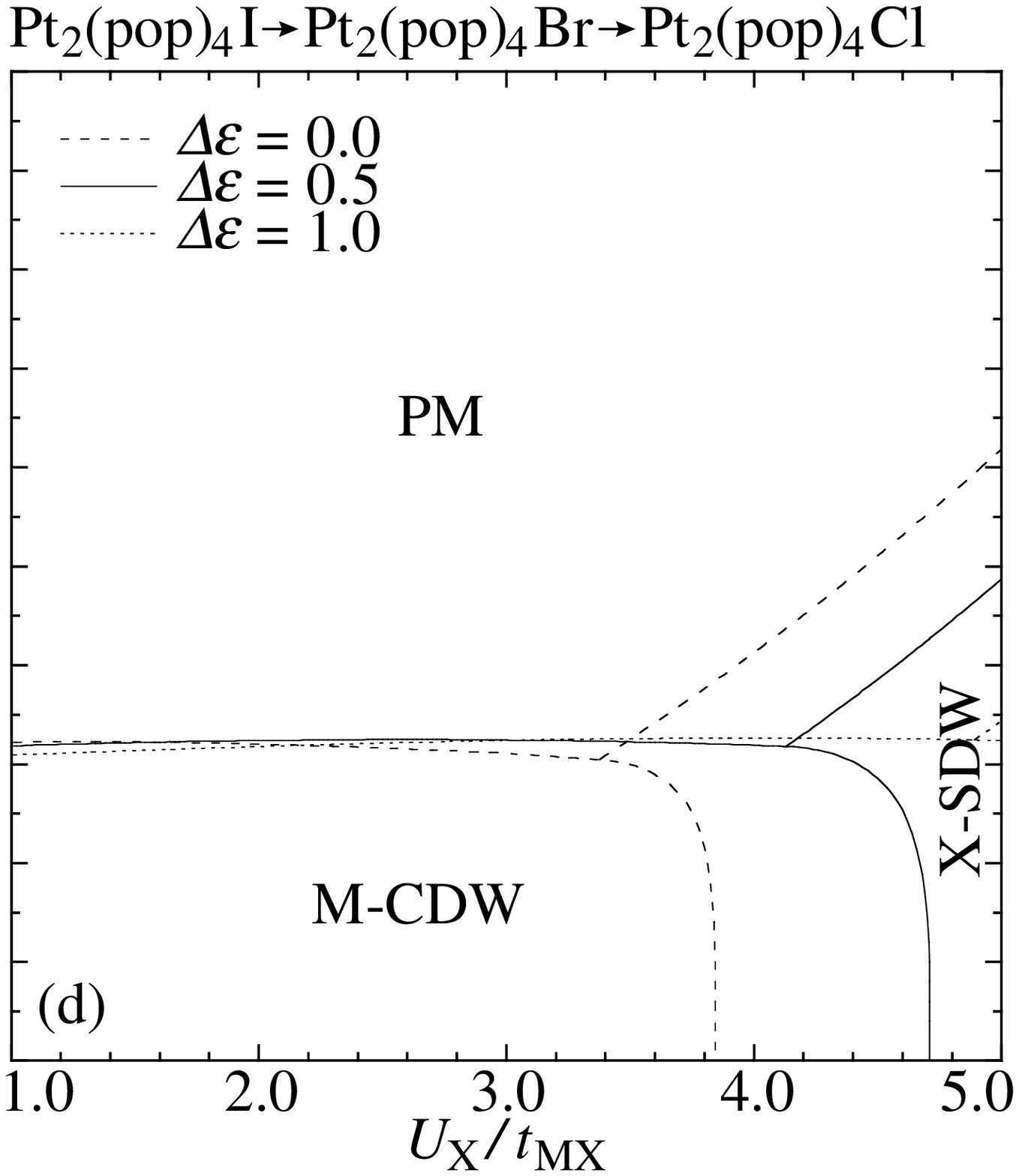,width=43.1mm,angle=0}}
\end{flushleft}
\vskip -22mm
\caption{Thermal phase boundaries as functions of
         the transfer integral
         (a: $\alpha=0.6$, $\beta=1.5$,
             $U_{\rm M}=2.0$, $U_{\rm X}=1.0$,
             $V_{\rm MM}=V_{\rm MX}=0.5$),
         the electron-lattice coupling
         (b: $t_{\rm MM}=1.2$, $\beta=1.5$,
             $U_{\rm M}=2.0$, $U_{\rm X}=1.0$,
             $V_{\rm MM}=V_{\rm MX}=0.5$),
         the M-site Coulomb repulsion
         (c: $t_{\rm MM}=1.2$, $\alpha=0.8$, $\beta=1.5$,
             $U_{\rm X}=1.0$, $V_{\rm MM}=V_{\rm MX}=0.5$)
         and the X-site Coulomb repulsion
         (d: $t_{\rm MM}=1.2$, $\alpha=0.6$, $\beta=1.5$,
             $U_{\rm M}=2.0$, $V_{\rm MM}=V_{\rm MX}=0.5$)
         at various values of
         ${\mit\Delta}\varepsilon\equiv
          \varepsilon_{\rm M}-\varepsilon_{\rm X}$.}
\label{F:PhD}
\end{figure}
\begin{figure}
\begin{flushleft}
\mbox{\qquad\qquad\quad\psfig{figure=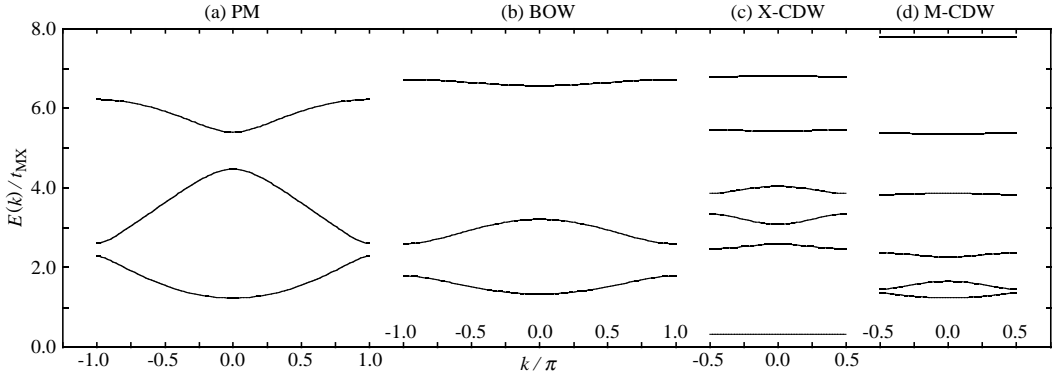,width=140.0mm,angle=0}}
\end{flushleft}
\vskip 0mm
\caption{Band dispersions of PM (a), BOW (b), X-CDW (c) and M-CDW (d)
         at $t_{\rm MM}=1.2$, $\alpha=0.8$, $\beta=1.5$,
         ${\mit\Delta}\varepsilon=0.5$,
         $U_{\rm M}=2.0$, $U_{\rm X}=1.0$ and
         $V_{\rm MM}=V_{\rm MX}=0.5$.}
\label{F:BD}
\end{figure}
\narrowtext

   We calculate the free energy for the broken-symmetry solutions
(a) to (h) of the Hamiltonian (\ref{E:H}) within the Hartree-Fock
approximation.
The lattice distortion is adiabatically treated and can therefore be
described as a function of the electron density matrices.
The numerical convergence of each solution is ensured and accelerated
by the irreducible decomposition of the mean-field Hamiltonian
\cite{SY}.
Surrounded with plenty of parameters, we observe the phase
competition especially from the point of view of electron itinerancy,
electron-phonon interaction and electron-electron correlation.

   Figure \ref{F:PhD}(a) is a good interpretation of the observed
thermal phase transitions.
With increasing temperature, X-CDW is transformed into PM via BOW,
whereas M-CDW directly into PM, each of which coincides with the
thermal behaviors of the dta and pop complexes, respectively.
The thermal stabilization of PM and BOW is convincing when we consider
their electronic structures.
In comparison with X-CDW and M-CDW with cell doubling, PM and BOW
maintain the original lattice symmetry and therefore possess
effectively half-filled conduction bands, which is visualized in Fig.
\ref{F:BD}.
Since an energy gap crossing over the Fermi level is disadvantageous
in gaining entropy via thermal excitations, PM and BOW may
compete with X-CDW and M-CDW at finite temperatures.
Figure \ref{F:BD} suggests that the conduction band for BOW is less
dispersive than that for PM, which can justify BOW preceding PM in
thermal stabilization.
The quantum fluctuation between X-CDW and M-CDW is also convincing
from the point of view of electron transfer.
The orbital hybridization within every M$_2$ moiety is essential in
the strongly {\it valence-trapped} M-CDW state, while it is the
overlap of the $d_{\sigma^*}$ orbitals on neighboring M$_2$ moieties
that stabilizes the relatively {\it valence-delocalized} X-CDW state.
Therefore, M-CDW is more stabilized with increasing $t_{\rm MM}$.
Based on semiempirical quantum-chemical band calculations, Borshch
{\it et al.} \cite{SB562} pointed out a possibility of twisting of
the dta ligand reducing the electronic communication by $\pi$
delocalization between adjacent metal sites and ending in X-CDW with
charge disproportionation in the M$_2$ moiety.
The transition from PM to BOW can also be understood as charge
disproportionation within every M$_2$ moiety with decreasing
$t_{\rm MM}$.
However, the contrast between the pop and dta complexes may not be
attributable mainly to their transfer integrals.
Early x-ray structural investigations \cite{RC409,CB444} reported the
Pt-Pt distance in the dimer to be $2.813$ \AA\ and $2.677$ \AA\ for
K$_4$[Pt$_2$(pop)$_4$Cl]$\cdot$$3$H$_2$O and Pt$_2$(dta)$_4$I,
respectively.
Even if we assume the general tendency
$t_{\mbox{Pt-Cl}}<t_{\mbox{Pt-Br}}<t_{\mbox{Pt-I}}$, it is rather
hard to conclude that $t_{\rm MM}/t_{\rm MX}$ for the dta complex is
generally smaller than those for the pop complexes.
The idea of larger M($d_{z^2}$)-M($d_{z^2}$) overlap of the dta
complex can further be justified by its higher conductivity
($13$ $\Omega^{-1}\,\mbox{cm}^{-1}$) compared with those of the
pop-family iodo complexes
($\sim 10^{-2}-10^{-4}$ $\Omega^{-1}\,\mbox{cm}^{-1}$).

   Then, what is the leading factor of such a striking contrast
between the pop and dta complexes?
We seek it in their electron-lattice interactions.
Figure \ref{F:PhD}(b) shows that M-CDW is stabilized with increasing
$\beta$, whereas X-CDW with increasing $\alpha$.
The orbital hybridization mainly stays within every M$_2$ moiety for
M-CDW, while it essentially extends over neighboring M$_2$ moieties
for X-CDW.
Therefore, X-CDW is quite sensitive to $\alpha$ effectively
describing the alternation of the MXM interdimer transfer.
In comparison with $\beta$ which is a measure for the mobility of the
X sublattice, $\alpha$ relatively describes the mobility of the
M$_2$ sublattice.
The halogen ions are rather free in both pop and dta complexes and
therefore $\beta$ may not characterize these materials well.
On the other hand, $\alpha$ must distinguish Pt$_2$(dta)$_4$I from
the pop-family compounds.
The M$_2$ moieties are tightly locked together in the pop complexes
due to the hydrogen bonds between the ligands and the counter
cations, whereas they are rather movable in the dta complex owing to
its neutral chain structure.
Thus, explicitly considering interdimer elastic constants as well, a
significantly larger $\alpha$ is expected for Pt$_2$(dta)$_4$I.
Now Fig. \ref{F:PhD}(b) allows us to attribute the distinct thermal
behaviors of the pop and dta complexes mainly to their
electron-phonon interactions.
We never exclude out the effect of their transfer integrals as an
underlying driving force.
Phase transitions from this point of view may definitely be observed
under external pressure \cite{SC723,BS405,MS066}.
As the adjacent metals in the dimer are tightly locked to each other
by the surrounding ligands, an applied pressure enhances $t_{\rm MX}$
rather than $t_{\rm MM}$.

   In relation to replacing the metals and the halogens, it is also
interesting to observe the phase competition as a function of the
Coulomb interactions.
Such a tuning of the electronic state is visualized in Figs.
\ref{F:PhD}(c) and \ref{F:PhD}(d).
Pt can be replaced by Ni in the dta complexes.
Considering that the orbital energies of Ni and I are so close as to
be possibly reversed
($\varepsilon_{\rm M}-\varepsilon_{\rm X}
  \equiv{\mit\Delta}\varepsilon<0$) in the crystal and
the on-site $d$-$d$ Coulomb repulsion is considerably strong
for Ni \cite{HO438}, Fig. \ref{F:PhD}(c) well supports the
observations \cite{RI907} for Ni$_2$(dta)$_4$I$-$no Peierls
distortion and the Mott-insulating behavior exhibiting an
antiferromagnetic spin susceptibility.
On the other hand, as ${\mit\Delta}\varepsilon$ increases, M-CDW
begins to compete with X-CDW.
This is quite convincing from the point of view of orbital
hybridization.
We have already learnt in Fig. \ref{F:DW} that X-CDW should be
characterized by the charge density wave on the X sublattice.
When we add up the charge densities on the adjacent metal sites in
every M$_2$ moiety, they shall no more modulate for X-CDW but still
oscillate for M-CDW. 
Furthermore Fig. \ref{F:BD} shows that the largest gap opens between
the two highest-lying bands primarily of $d_{\sigma^*}$ character for
M-CDW, whereas between the two lowest-lying ones for X-CDW implying
an essential hybridization of the M $d_{z^2}$ and X $p_z$ orbitals.
We are thus convinced that M-CDW and X-CDW are stabilized and
destabilized, respectively, with the increase of
${\mit\Delta}\varepsilon$.
Figure \ref{F:PhD}(c) is then an interpretation of the close Pt
$d_{z^2}$ and I $p_z$ orbital energies.
Finally we take a look at Fig. \ref{F:PhD}(d) raising a possibility of
the novel antiferromagnetic state X-SDW appearing by the replacement
of the halogens.
$U_{\rm X}$ increases in close cooperation with the increase of
${\mit\Delta}\varepsilon$ and therefore its stabilization is not so
trivial in practice.
A somewhat tricky situation $V_{\rm MM}<V_{\rm MX}$ advantageously
acts on X-SDW \cite{SY}.
Otherwise we wonder whether a Pd chain chloride could be synthesized.

   We hope that the present calculations will settle the
controversial observations especially for Pt$_2$(dta)$_4$I
\cite{CB444,CB815,HK068} and contribute to a semiquantitative
estimation of the essential parameters.
In such circumstances that even the hopping amplitudes are not yet
revealed, a close collaboration between experimental and theoretical
investigations is really expected.

   The author is grateful to Prof. K. Yonemitsu and Dr. K. Iwano for
useful discussions.
He thanks Prof. H. Okamoto and Prof. H. Kitagawa as well for helpful
comments on MMX materials.
This work was supported by the Japanese Ministry of Education,
Science, and Culture.
The numerical calculation was done using the facility of the
Supercomputer Center, Institute for Solid State Physics, University
of Tokyo.

\widetext
\end{document}